\documentclass[letterpaper,aps,prb,preprint,superscriptaddress]{revtex4}

\usepackage{amsmath}
\usepackage{graphicx}
\usepackage[colorlinks=true,citecolor=blue,linkcolor=blue,urlcolor=blue]{hyperref} 
\usepackage{float}
\usepackage{txfonts}
\usepackage{bm}
\usepackage{xspace}
\usepackage{units}
\usepackage{array}
\usepackage{tabulary}

\begin{document}

\definecolor{delftdark}{rgb}{0, .4, .6}
\newcommand{\todo}[1]{\textcolor{delftdark}{!![\,#1\,]!!}}

\begin{centering}
\textbf{\large Benchmark and application of unsupervised classification approaches for univariate data}\\[0.5cm]

Maria El Abbassi$^1$, Jan Overbeck$^{2,3,4}$, Oliver Braun$^{2,3}$, Michel Calame$^{2,3,4}$, Herre S.J. van der Zant$^1$ and Mickael L. Perrin$^{2,*}$ \\
\textit{$^1$ Kavli Institute of Nanoscience, Delft University of Technology, Lorentzweg 1, 2628 CJ Delft, The Netherlands}\\
\textit{$^2$ Empa, Swiss Federal Laboratories for Materials Science and Technology, \"{U}berlandstrasse 129, CH-8600 D\"ubendorf, Switzerland.}\\
\textit{$^3$ Department of Physics, University of Basel, Klingelbergstrasse 82, CH-4056 Basel, Switzerland.}\\
\textit{$^4$ Swiss Nanoscience Institute, University of Basel, Klingelbergstrasse 82, CH-4056 Basel, Switzerland.}\\
\textit{$^*$ email: Mickael.Perrin@empa.ch.}\\[1.0cm]
\end{centering}

\textbf{Abstract}: Unsupervised machine learning, and in particular data clustering, is a powerful approach for the analysis of datasets and identification of characteristic features occurring throughout a dataset. It is gaining popularity across scientific disciplines and is particularly useful for applications without a priori knowledge of the data structure. Here, we introduce an approach for unsupervised data classification of any dataset consisting of a series of univariate measurements. It is therefore ideally suited for a wide range of measurement types. We apply it to the field of nanoelectronics and spectroscopy to identify meaningful structures in data sets. We also provide guidelines for the estimation of the optimum number of clusters. In addition, we have performed an extensive benchmark of novel and existing machine learning approaches and observe significant performance differences. Careful selection of the feature space construction method and clustering algorithms for a specific measurement type can therefore greatly improve classification accuracies.

\section*{INTRODUCTION}
Machine learning (ML) and artificial intelligence are among the most significant recent technological advancements, with currently billions of dollars being invested in this emerging technology\cite{AI_spending2020}. In a few years, complex problems which had been around for decades, such as image\cite{Schmidhuber2015} and facial recognition\cite{Sun2014, Liu2015}, speech\cite{Mikolov2010,Hinton2012} and text\cite{Zhang2015,Tshitoyan2019} understanding, have been addressed. Machine learning promises to be a game-changer for major industries like health care\cite{Kourou2015}, pharmaceuticals\cite{Vamathevan2019}, information technology\cite{Hutto2014}, automotive\cite{Bojarski2016} and other industries relying on big data\cite{Chen2014}. Its underlying strength is the excellence at recognizing patterns, either by relying on previous experience (supervised ML), or without any a priori knowledge of the system itself (unsupervised ML). In both cases, ML relies on large amounts of data, which, in the last two decades, have become increasingly available due to the fast rise of cheap consumer electronics and the internet of things.

The same trend is also observed for scientific research, including the field of nanoscience, where tremendous progress has been made in the data acquisition\cite{Graf2007,ElAbbassi2019c,Brown2020} and public databases have become available containing, for instance, a vast number of material structures and properties\cite{Curtarolo2013,Pizzi2016}. Inspiring examples of the use of the predictive power of supervised machine learning have, for instance, been realized in quantum chemistry for the prediction of the quantum mechanical wave function of electrons\cite{Schutt2019} and in nanoelectronics for the tuning of quantum dots\cite{Lennon2019}, the identification of 2D material samples\cite{Masubuchi2020}, and the classification of breaking traces in atomic contacts\cite{Lauritzen2018}. Unsupervised machine learning methods, on the other hand, are intended for the investigation of the underlying structure of datasets without any a priori knowledge of the system. Such approaches are ideally suited for the analysis of large experimental datasets and can help to significantly reduce the issue of conformation bias in the data analysis\cite{Ioannidis2005}. 

Several studies involving data clustering in nanoelectronics applications have been reported to date\cite{Lemmer2016,Wu2017,Hamill2018,Cabosart2019,ElAbbassi2019,Huang2019,Vladyka2020,Bamberger2020}. 
In the study by Lemmer et al.\cite{Lemmer2016}, the univariate measurement data (conductance versus electrode displacement) is treated as an $M$-dimensional vector and compared to a reference vector for the feature space construction, after which the Gustafson-Kessel (GK) algorithm\cite{Gustafson1978} is employed for classification. A variation of this method was applied by El Abbassi et al.\cite{ElAbbassi2019} to current-voltage characteristics. In a more recent study\cite{Cabosart2019}, the need for this reference vector was eliminated by creating a 28$\times$28 image of each measurement trace. However, the high number of dimensions resulting from this approach is problematic for many clustering algorithms, as the data becomes sparse for increasing dimensionality (curse of dimensionality\cite{Bellman2010}), thereby restricting the available clustering algorithms. Several approaches have been proposed to reduce the number of dimensions, such as deep auto-encoder for feature extraction from the raw data itself\cite{Huang2019}, or the use of the approximately linear sections of the breaking traces\cite{Bamberger2020}. Characteristic of the previous studies, however, is the fact that the clustering is performed on a feature space constructed from the individual breaking traces, an approach that can become computationally prohibitive in case large datasets are acquired. An appealing alternative has been introduced by Wu el al.\cite{Wu2017}, in which the clustering algorithms is run on the 2D conductance-displacement histogram. 

In all above-mentioned studies, only a single feature space construction method and clustering algorithm were investigated, without a systematic benchmark of their accuracy against a large number of datasets of known classes and with varying partitions. This makes it difficult to compare the performance of one method to another. In addition, few studies\cite{Wu2017,Bamberger2020} provide guidelines for the estimation of the number of clusters, a critical step in data partitioning.

Here, we provide a workflow for the classification of univariate data sets. Our three-step approach consists of 1. the feature space construction, 2. the clustering algorithm, and 3. the internal validation to define the optimum number of clusters (NoC). In the first part of the article, we benchmark a wide range of 28 feature space construction methods as well as 16 clustering algorithms using 900 datasets of simulated breaking traces with a number of classes varying between 2 and 10. In this benchmark, we identify the top five best performing clustering algorithms and top two feature spaces. We then apply our workflow to several distinctively different measurement types (break-junction conductance traces, current-voltage characteristics, and Raman spectra), yielding extracted clusters that are distinctively different. Importantly, our approach does not require any a priori knowledge of the system under study and therefore reduces the confirmation bias that may be present in the analysis of large scientific datasets. The attribution of the various clusters to the physical phenomena dictating their behaviour, however, requires a detailed understanding of the microscopic picture of the system under study and is beyond the scope of this article. \\

\section*{RESULTS}

\begin{figure}
\centering
\includegraphics[width=0.99\textwidth]{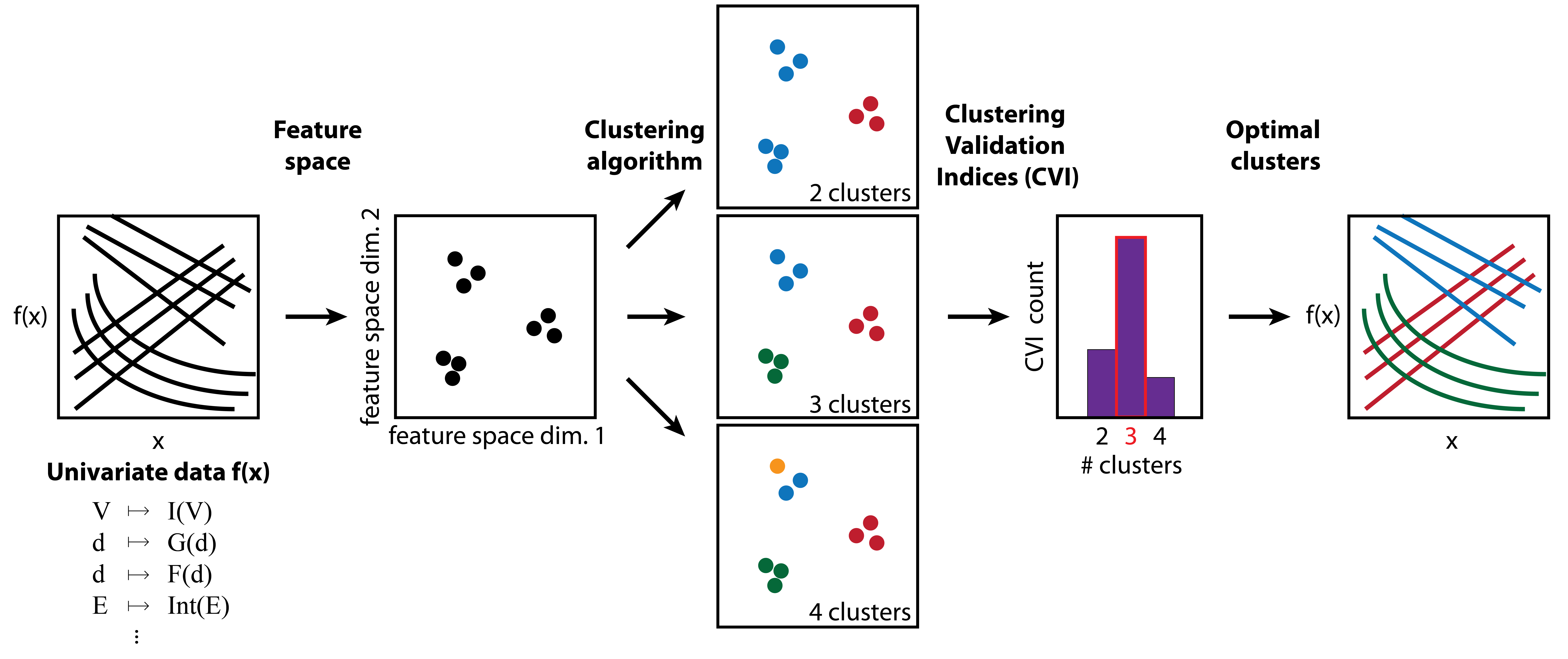} 
\caption{\label{fig:intro} \textbf{Concept of our approach for univariate data classification.} Any dataset in which the data depends on a single variable ( for instance current $I$ vs. bias voltage $V$, conductance $G$ vs. electrode displacement $d$, force $F$ vs. displacement $d$, intensity $Int$ vs. energy $E$, etc... ) can be converted into a feature vector. The feature space spanning the entire dataset is then split into clusters (represented using different colors) using a clustering algorithm. Finally, cluster validation indices (CVIs) are used to estimate the optimal number of clusters (NoC). }
\end{figure}

A schematic of the workflow for the unsupervised classification of univariate measurements is depicted in Fig.~\ref{fig:intro}, starting from a dataset consisting of $N$ univariate and discrete functions $f(x_i)$, i $\in$ [1,N]. Each measurement curve is converted into an $M$-dimensional feature vector, resulting in a feature space containing $M \times N$ data points. After this step, a clustering algorithm is applied. As the number of classes is not known a priori, this clustering step is repeated for a range of cluster numbers (in this illustration for 2-4 clusters). 
Here, we define a class as the ground truth distribution of each dataset, and a cluster the result of a clustering algorithm. Then, in order to determine the most suited NoC and assess the quality of the partitioning of the data, up to 29 internal cluster validation indices (CVIs) are employed. Each CVI provides a prediction for the NoC, after which the optimal NoC is estimated based on a histogram of the predictions obtained from all CVIs. These CVIs are also used to determine the optimal feature space method and clustering algorithm.\\

\subsection*{Benchmarking of algorithm performance on simulated mechanically controllable break-junction datasets}

In the following, a large variety of feature space construction methods and clustering algorithms are investigated and their performance benchmarked against artificially created datasets with known classes. The aim of a benchmark is to rank the various algorithms according to their performance for a given set of parameters. Here, all algorithms were executed using their default parameters, both in the benchmark, as well as when applied to experimental datasets.
The simulated datasets are conductance-displacement traces - also known as breaking traces - as commonly measured using the mechanically controllable break-junction (MCBJ) technique and scanning tunneling microscrope (STM) for measuring the conductance of a molecule\cite{Xu2003}. For a detailed description of the construction of the simulated (labeled) data, we refer to Supplementary Method 1. 

In short, we generated 900 datasets, each consisting of 2000 breaking traces with known labels, with a varying number of classes between 2 and 10 (100 x 2 classes ... 100 x 10 classes). The traces were generated based on an experimental dataset consisting of conductance vs. distance curves recorded on OPE3 molecules\cite{Frisenda2018,Data_experimental}. This is in contrast to previous studies where the benchmark data was purely synthetic\cite{Lemmer2016,Huang2019}. To account for possibly large variations in cluster population which may occur experimentally, the distribution of classes is logarithmically distributed with the most probable class having 10 times more traces than the least occurring one. For example, for 2 classes the distribution is 9.09\% and 90.91\%, for 3 classes 6.10\%, 33.35\%, and 60.55\%, etc.... 

We applied a variety of feature space construction processes and clustering algorithms to each of these 900 datasets. We investigated vector-based feature space construction methods based on a reference vector as described in Lemmer et al.\cite{Lemmer2016}, feature extraction from the raw data itself\cite{Huang2019}, and conversion to images (two-dimensional histogram)\cite{Cabosart2019}. In the latter case, inspired by the MNIST datasets\cite{MNIST}, measurements are converted into images of 28$\times$28 pixels. This has the advantage that all inputs for the feature space construction method have the same size, independent on the number of data points in each measurement. 
Here, we would like to stress that the number of pixels can be chosen to fine-tune the resolution for the feature extraction, independently from the number of data points in the measurements. In Supplementary Note 1, we show that 28$\times$28, inspired by the MNIST database, is a good compromise between accuracy and computational cost. This choice implies that the distinction between features occurring below the bin size (0.25 orders of magnitude in conductance and 0.1~nm in distance) is limited as it relies only on the counts within the bin itself. To illustrate this, for fixed acquisition rate, a slanted plateau can be separated from a horizontal plateau as both would yield different counts in a particular bin. For a distinction between more elaborate shapes a denser grid would be beneficial. However, the use of more bins comes at higher computational costs and may lead to high-dimensional sparse data, which in turn is challenging to cluster, even after dimensionality reduction.

In the following, the three different approaches will be referred to as `Lemmer', `raw', and `28$\times$28'. The high number of dimensions for the raw and 28$\times$28 case is known to lead to the curse of dimensionality\cite{Bellman2010}; the data becomes highly sparse and causes severe problems for many common clustering algorithms. To avoid this limitation, we have investigated a range of dimensionality reduction techniques, such as principal component analysis\cite{vandermaaten2009} (PCA), kernel-PCA\cite{vandermaaten2009}, multi-dimensional scaling\cite{vandermaaten2009} (MDS), deep autoencoders\cite{vandermaaten2009} (AE), Sammon mapping\cite{Sammon1969}, stochastic neighbor embedding\cite{Hinton2002} (SNE), t-distributed SNE\cite{VanDerMaaten2008} and uniform manifold approximation and projection\cite{McInnes2018} (UMAP). For the last two methods, three distance measure approaches were used (Euclidean, Chebyshev and cosine, abbreviated as Eucl., Cheb. and cos., respectively), bringing the total number of feature space construction methods to 28. For all methods containing dimensionality reduction, we used a reduction down to 3 dimensions. A description of each method is presented in Supplementary Method 2. In Supplementary Note 2, we show that by increasing the dimensions for t-SNE (cos.) from 3 to 7 only a marginal gain in Fowlkes-Mallows index can be achieved for the five selected algorithms.\\

\begin{figure}
\centering
\includegraphics[width=0.99\textwidth]{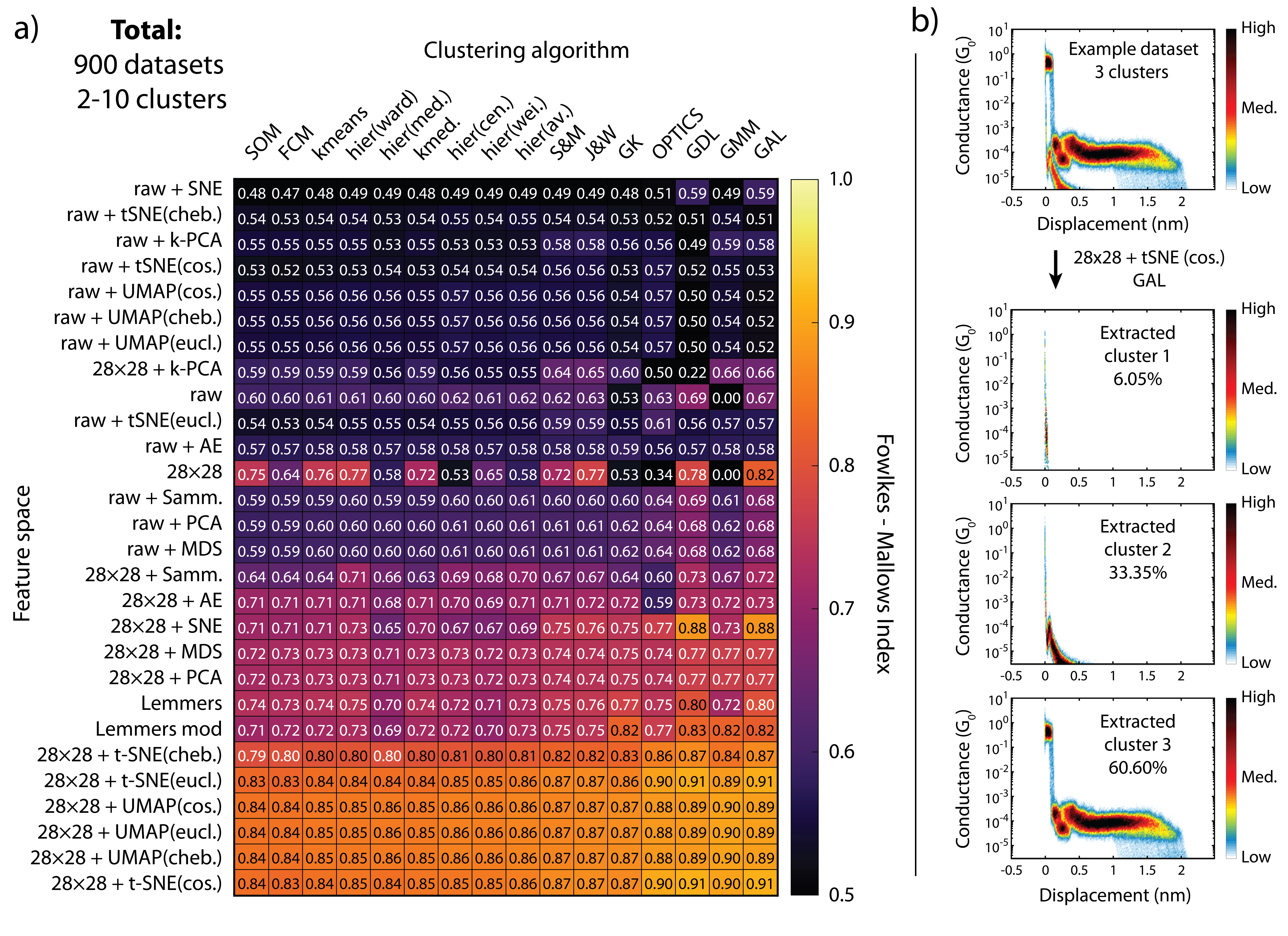} 
\caption{\label{fig:benchmark} \textbf{Benchmarking of various feature spaces and clustering algorithm on simulated mechanically controllable break-junction data.} a) Overview of the accuracy, expressed as Folwkes-Mallows (FM) index, for all combinations of the various feature space construction methods and clustering algorithms. For this analysis, the average FM is shown based on 900 datasets of 2000 traces each, with 2-10 classes. The rows and columns of the heatmap have been sorted by increasing average FM-index, with the best combination of feature spaces and algorithm in the lower right corner. b) 2D conductance-displacement histogram for an example dataset, including the 2D conductance-displacement histograms obtained by clustering using the best performing feature space method 28$\times$28 + t-distributed stochastic neighbor embedding (t-SNE) using a cosine distance (cos.) and the graph average linkage (GAL) clustering method. }
\end{figure}

After each of the 900 datasets was run through the 28 feature space construction methods, 16 clustering algorithms were tested, covering a large spectrum of classification methods such as distance minimization methods (k-means, k-medoids), fuzzy methods (fuzzy C-mean\cite{Bezdek1981} (FCM) and GK\cite{Gustafson1978}), self-organizing maps\cite{Kohonen1982} (SOM), hierarchical methods\cite{Silla2011} with various distance measures, expectation-maximization methods (Gaussian mixed model\cite{Williams1995} (GMM)), graph-based agglomerative methods (graph degree linkage\cite{Zhang2012} (GDL) and graph average linkage\cite{Zhang2013} (GAL)), spectral methods (Shi and Malik\cite{Shi2000} (S\&M) and Jordan and Weiss\cite{Ng2001} (J\&W)) and density-based methods (Ordering Points To Identify the Clustering Structure (OPTICS\cite{Ankerst1999})). A description of each method can be found in Supplementary Method 3. We note that we restricted ourselves to algorithms in which the number of clusters can be explicitly defined as input parameter. This step is needed further on to calculate the data partitioning for 2 to 9 clusters and determine the optimum number of clusters using clustering validation indices.  
This restriction excludes algorithms such as DBSCAN\cite{Ester1996} (density-based spatial clustering of applications with noise), hierarchical DBSCAN (HDBSCAN\cite{Campello2013}), and affinity propagation\cite{Frey2007}. We also note that many different image classification algorithms are available that can be run directly on the 28$\times$28 image before dimensionality reduction, such as Deep Adaptive image Clustering\cite{Chang2017} (DAC), Associative Deep Clustering\cite{Haeusser2019} (ADC) and Invariant Information Clustering\cite{Ji2018} (IIC). Most of these algorithms, however, are based on neural networks and are significantly more expensive in terms of computational cost, thus limiting their applicability. The execution speeds of the various feature space and clustering methods applied here is presented in Supplementary Note 3.

The accuracy of the classification is evaluated using the Fowlkes-Mallows (FM) index\cite{Fowlkes1983}; it is an external cluster validation index (CVI) which varies between 0 and 1, where 1 represents the case of clusters perfectly reproducing the original classes. The FM index is defined as $ FM = \sqrt{ \frac {TP}{TP+FP}\cdot {\frac {TP}{TP+FN}}}$, where $TP$ is the number of true positives, $FP$ is the number of false positives, and $FN$ is the number of false negatives. 
The mean Fowlkes-Mallows indices for all combinations of feature space and clustering approach based on all 900 datasets are shown in Fig.~\ref{fig:benchmark}a, presented as heatmap.  Figure~\ref{fig:benchmark}b presents an example dataset that has been clustered using t-SNE (cos.) and GAL. We note that the NoC used for clustering is chosen to be the same number as the number of classes provided in the simulated dataset. The heatmap is sorted by increasing average FM index per column and row, respectively, with the most accurate combination in the lower right corner. In this extensive benchmark, the least accurate algorithm is raw + SNE combined with FCM with a FM index of 0.47, while the most accurate one is the 28$\times$28 + t-SNE (cos.) feature space, combined with the GAL algorithm. Based on the benchmark performed on this dataset, this optimal combination feature space and clustering algorithm exhibits a FM index of 0.91 and outperforms previously used methods to classify similar datasets in literature\cite{Lemmer2016,Cabosart2019,Huang2019}. 

The heatmap also shows that both 28$\times$28 + t-SNE and 28$\times$28 + UMAP perform similarly well and provide a significant improvement in accuracy with respect to the other feature space methods investigated. 
In the following, we will therefore focus on these two feature space methods using the cosine distance measure.
In terms of the clustering algorithm, the heatmap shows that the GAL algorithm yields the highest accuracy. This observation follows a previous study demonstrating that GAL outperforms many state-of-the-arts algorithms for image clustering and object matching\cite{Zhang2012}. 

To ensure that the benchmark is not biased by the use of a logarithmically distributed class population, we produced the same heatmap as shown in Fig.~\ref{fig:benchmark}a but on datasets containing equal-size classes (see Supplementary Note 4). This benchmark yields very similar results in terms of best performing feature spaces and clustering algorithms. Finally, to account for different noises that may be present during experiments, we generated three additional datasets (see Supplementary Note 4 for details). One dataset had an increased amount of noise, while the two others contained heteroscedastic noise, either scaling with conductance or with displacement. The best performing feature spaces and clustering algorithms remain largely unaffected.

From the fact that the row-to-row variation of FM indices, i.e., between feature space methods, is larger than the difference between columns (clustering methods), we conclude that the role of the feature space is more important than that of the algorithm. This can be rationalized, as a better feature space method will produce distinctively separated clusters, making it easier for the algorithm to find these clusters. However, as this benchmark is performed on synthetic data, the performance of the algorithms may be different than on actual data. Therefore, we select the five best performing algorithms, namely GK, the most accurate of the spectral methods (J\&W), GMM, the most accurate graph-based method (GAL), and OPTICS for further studies in the remainder of this paper. 

\section*{Application to an experimental mechanically controllable break-junction dataset}
\begin{figure}[t]
\centering
\includegraphics[width=0.99\textwidth]{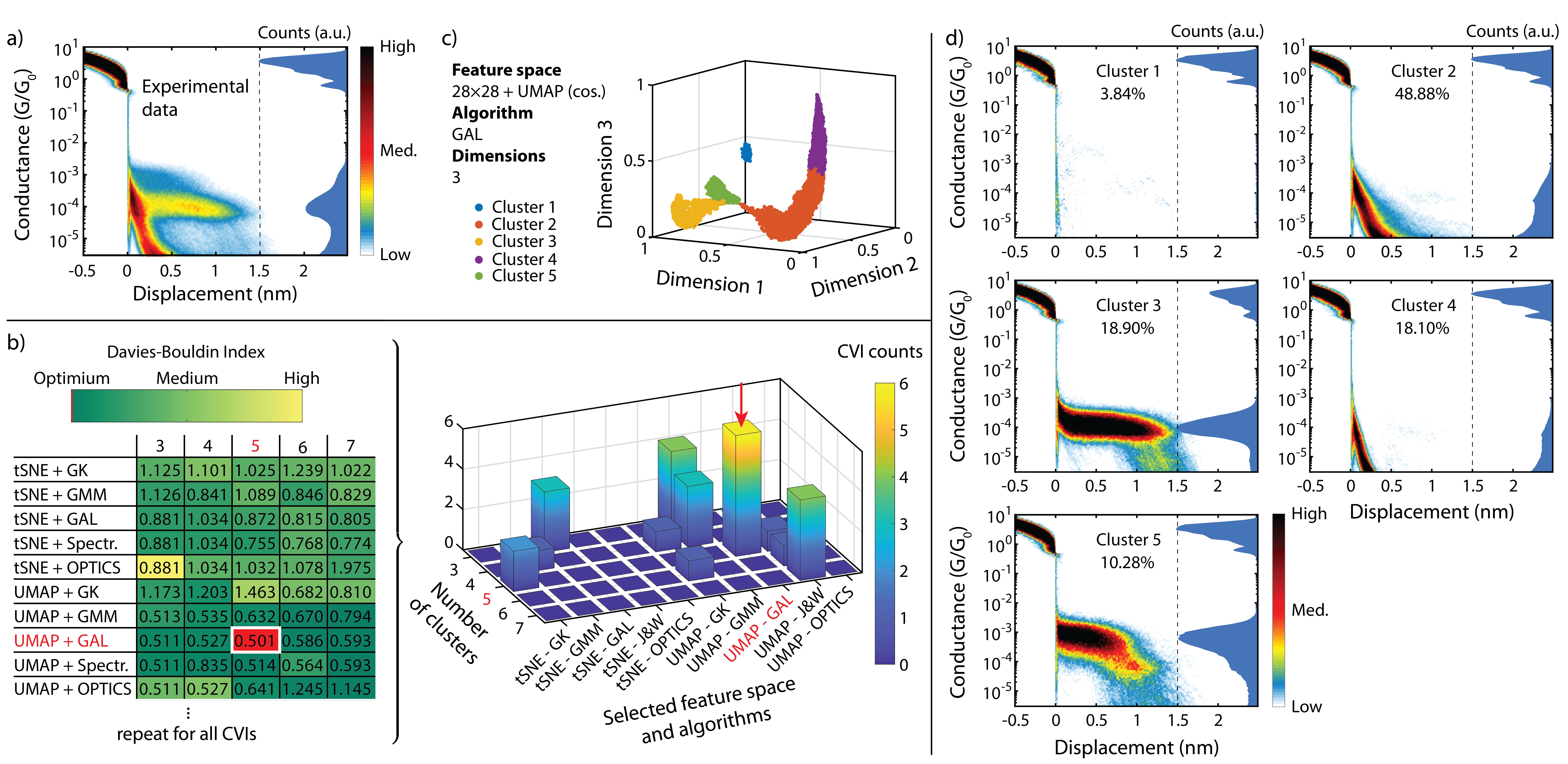} 
\caption{\label{fig:MCBJ} \textbf{Application of the workflow to measured mechanically controllable break-junction data.} a) Experimental 2D conductance-displacement histogram based on 10'000 breaking traces. The blue area represents the corresponding 1D conductance histogram. b) Determination of the most suited feature space, clustering algorithm and the optimal number of clusters using cluster validation indices (CVI). The feature space considered are 28$\times$28 + t-distributed stochastic neighbor embedding (t-SNE) and 28$\times$28 + uniform manifold approximation and projection (UMAP), both using a cosine (cos.) distance metric. The clustering algorithms considered are Gustafson-Kessel (GK), Gaussian mixed model (GMM), graph average linkage (GAL), spectral clustering following Jordan and Weiss (J\&W) and Ordering Points To Identify the Clustering Structure (OPTICS). The heat map represents the Davies-Bouldin CVI, requiring a minimization of its value (red/white highlighted box). The histogram counts the occurrence of the feature space and clustering algorithm combinations, and of the optimal number of clusters, as predicted by the various CVIs. c) Feature space constructed from the data of (a) using 28$\times$28 + UMAP (cos.) and the GAL clustering method for 5 clusters. d) 1D conductance and 2D conductance-displacement histogram for the cluster assignment in (c). }
\end{figure}

We now apply our workflow to an experimental dataset of unknown classes and illustrate the different steps in Fig.~\ref{fig:MCBJ}. The starting point is an MCBJ dataset consisting of 10'000 traces recorded on the OPE3 molecule\cite{Frisenda2018} (see Fig.~\ref{fig:MCBJ}a for the 2D conductance-displacement histogram), to which we apply the two selected feature space methods 28$\times$28 + t-SNE (cos.) and 28$\times$28 + UMAP (cos.). Subsequently, these feature spaces are classified using the five selected clustering methods for a NoC ranging from 2 to 8. This gives a total of 5x2x7 = 70 different clustering distributions. For each of them, we calculate internal cluster validation indices\cite{Arbelaitz2013,Charrad2014,Hamalainen2017,NBCluster} (CVIs). Each index is calculated for a varying NoC, from which the optimum NoC can be estimated by different means (minimum, maximum, elbow, etc...). 
Here, we choose 29 CVIs, including the well-known Silhouette index, Dunn and Davies-Bouldin index, that only require a maximization/minimization of the index. As such, the index can be used to compare different clustering methods, feature space, and NoCs, and determine the optimum combination. A complete list of all the indices and their implementation can be found in Supplementary Method 4.

The heat map shown in Fig.~\ref{fig:MCBJ}b presents the calculated values of the Davies-Bouldin index as a matrix, with as columns the NoC and as rows all combinations of feature space and the clustering algorithm. From this matrix, the maximum/minimum value of the index is obtained to determine the optimum NoC and method as determined by this particular CVI. We note that the use of CVIs to estimate the NoC is not straightforward as each of them has implicit assumptions, in particular on the distribution of the clusters. For this reason, we only consider NoC estimations that are unambiguous, in other words, a well-defined peak or dip in the cluster validation index. This means that we calculate the CVIs for 2 to 8 clusters, but we only take the CVI into account if the optimum NoC lies between 3 to 7 clusters. This procedure is repeated for all 29 CVIs and a 2D histogram is constructed (Fig.~\ref{fig:MCBJ}b). Finally, this allows us to directly access the overall best feature space (28$\times$28 + UMAP), algorithm (GAL) and NoC (5). As a verification of the robustness of the CVI prediction, we have performed the same analysis including, in addition, two poorly performing feature spaces (raw + t-SNE (cos.), and raw + UMAP (cos.)) and the same five clustering algorithms.
Shown in Supplementary Note 5, the analysis shows that the combination of 28$\times$28 + UMAP, GAL and 5 clusters again comes out as optimal. The resulting feature space, with the individual breaking traces colored by cluster assignment, is plotted in Fig.~\ref{fig:MCBJ}c.

The resulting clusters are visualized as 2D conductance displacement histograms built from the individual breaking traces (see Fig.\ref{fig:MCBJ}d). The plots show that the resulting 2D histograms exhibit distinctively different breaking behaviors, and based on our knowledge of these junctions, one can speculate that Cluster 1 corresponds to gold junctions breaking directly to below the noise floor, Cluster 2 to tunneling traces with some hints of molecular signatures, Cluster 3 to a fully stretched OPE3 molecule, Cluster 4 to tunneling traces without any molecular presence, and Cluster 5 to a two step breaking process involving molecule-electrode interactions. The exact attribution of the various clusters, however, requires a detailed understanding of the microscopic picture of the molecular junction, possibly supported by ab-initio calculations, and is beyond the scope of this article. 
Even though the CVIs show that five is the ideal number of clusters, this result should be taken with a grain of salt. To the best of our knowledge, no CVI exists that performs well in all situations. In particular clusters of largely varying densities are challenging as well as clusters of arbitrary shape. Therefore, the CVIs should be used mere as a guideline, and, as reference, we show the resulting cluster for 3-7 clusters in Supplementary Note 6.

To illustrate the versatility of our approach for different measurements types, we now proceed with the classification of two more datasets: the first one consists of 67 current-voltage (IV) characteristics, while the second one contains 4900 Raman spectra. For the current-voltage characteristics classification, we note that the OPTICS algorithm was excluded as it fails using the default parameters due to the limited amount of measurements. 

\subsection*{Application to current-voltage characteristics}

\begin{figure}[ht]
\centering
\includegraphics[width=0.99\textwidth]{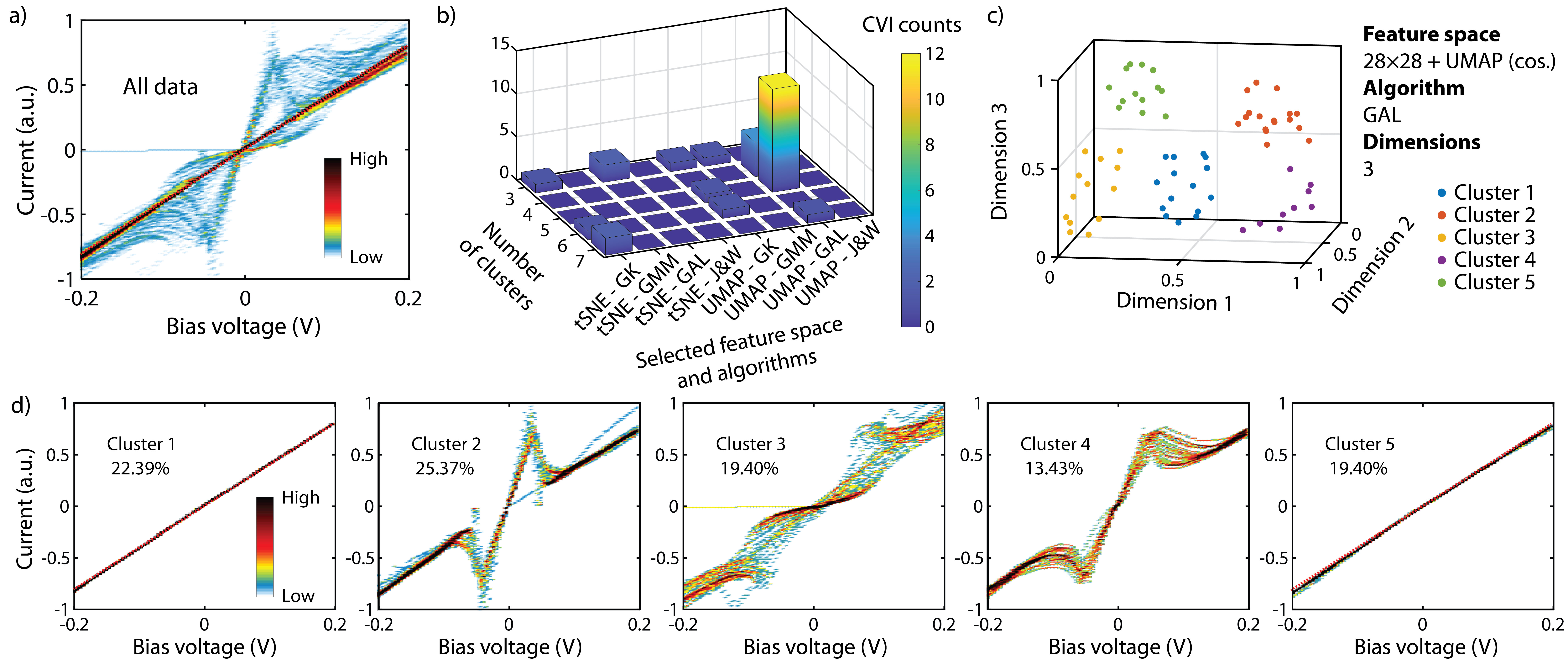} 
\caption{\label{fig:IVs} \textbf{Application of the method on current-voltage characteristics.} a) Experimental 2D current-voltage histogram based on 67 current-voltage characteristics recorded on a dihydroanthracene molecule\cite{Perrin2014,Perrin2020,Data_experimental}.
b) Determination of the most suited feature space, clustering algorithm and the optimal number of clusters using cluster validation indices (CVI). The feature space considered are 28$\times$28 + t-distributed stochastic neighbor embedding (t-SNE) and 28$\times$28 + uniform manifold approximation and projection (UMAP), both using a cosine (cos.) distance metric. The clustering algorithms considered are Gustafson-Kessel (GK), Gaussian mixed model (GMM), graph average linkage (GAL), spectral clustering following Jordan and Weiss (J\&W) and Ordering Points To Identify the Clustering Structure (OPTICS).
c) Feature space constructed using the 28$\times$28 + UMAP (cos.) and clustered using the GAL algorithm for 5 clusters.
d) 2D current-voltage histogram of the data shown in (a), clustered according to the partitioning shown in (c).
}\end{figure}

Figure~\ref{fig:IVs}a presents a 2D current-voltage histogram of 67 IV characteristics recorded on a dihydroanthracene molecule\cite{Perrin2014,Perrin2020,Data_experimental}. The IVs have been normalized to focus on the shape of the curves, not on the absolute values in current. The same clustering procedure is repeated as described previously and the best feature space and clustering algorithm is determined to be 28$\times$28 + UMAP(cos.) and GAL, respectively for an optimal number of clusters of 5 (Figure~\ref{fig:IVs}b). The corresponding feature space is presented in Fig.~\ref{fig:IVs}c, colored according to the clusters produced by the GAL algorithm. The 2D current-voltage histograms of the five resulting clusters are shown in Fig.~\ref{fig:IVs}d. Cluster 1 shows perfectly linear IVs, while cluster 2 shows a pronounced negative differential conductance (NDC) feature, with first a linear slope around zero bias, a sharp peak around 30~mV, followed by a rapid decrease of the current for increasing bias voltage. Cluster 3 contains mostly IVs with a gap around zero bias. Cluster 4 exhibits NDC as well, but with a more rounded peak compared to cluster 2, and a more gentle decrease in current. Cluster 5 shows close-to-linear IVs with some deviations from the perfect line. \\

\section*{Application to Raman spectra}
\label{sec:Raman}
As a final application, we investigate the classification of Raman spectra\cite{Data_experimental}. As Raman spectra are less stochastic than MCBJ measurements, we have performed a separate benchmark (see Supplementary Note 7) to rank the different algorithms. We find that, similar as for the MCBJs, the 28$\times$28 + t-SNE and 28$\times$28 + UMAP feature spaces perform the best. For the clustering algorithms, we find that most of them perform similarly well.

\begin{figure}[b]
\centering
\includegraphics[width=0.99\textwidth]{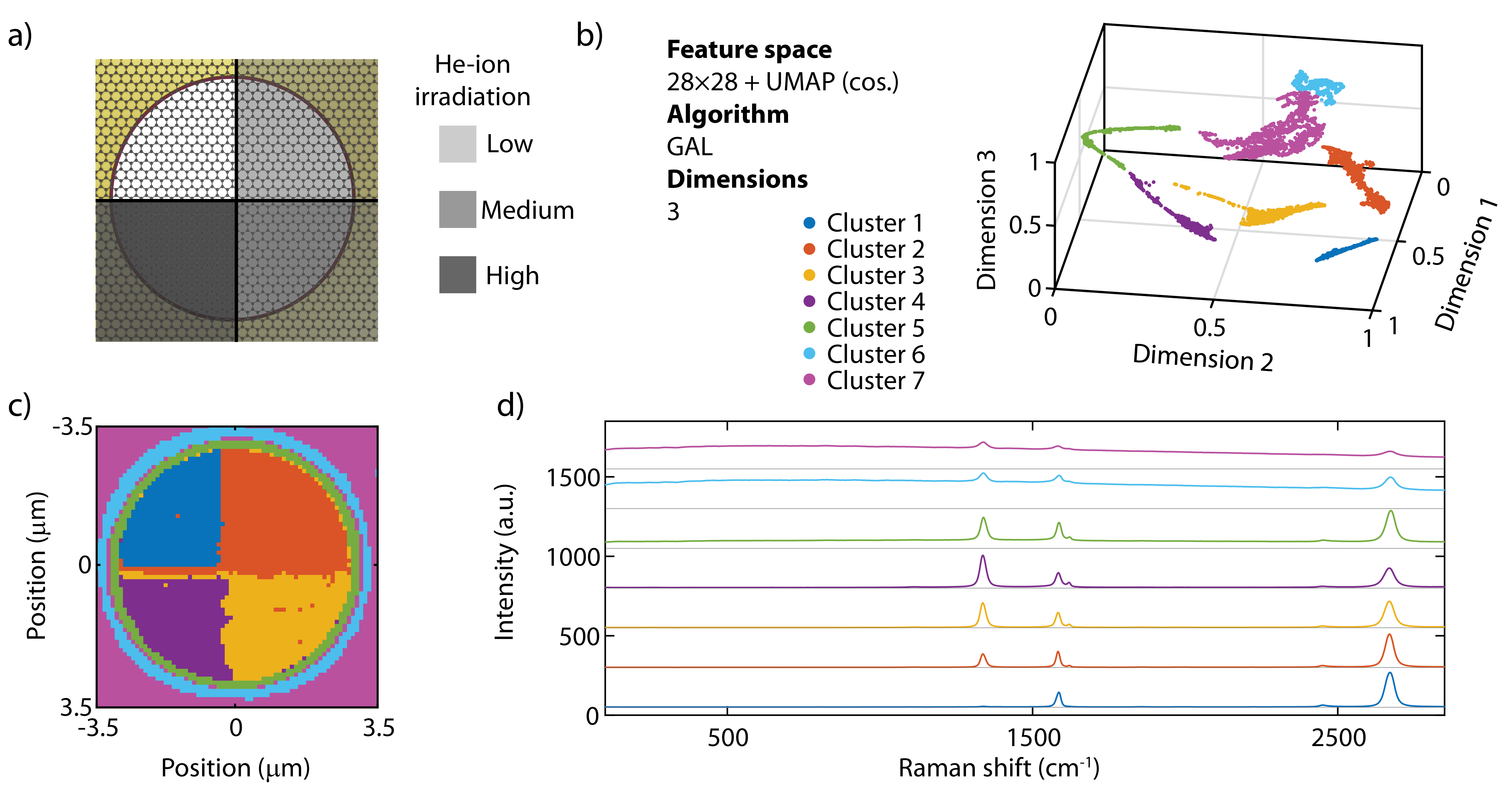} 
\caption{\label{fig:Raman} \textbf{Application of the method on Raman spectra.} 
a) Sample layout: suspended graphene membrane irradiated with four different He-ion doses.
b) Partitioned feature space, constructed with 28$\times$28 + uniform manifold approximation and projection (UMAP) using the cosine (cos.) distance metric and the graph average linkage (GAL) clustering algorithm. 
c) Spatial map of the extracted clusters.
d) Average Raman spectrum of each cluster.
}
\end{figure}

The Raman spectra are recorded on a well-studied reference system, namely a graphene membrane that has been divided in four quadrants, each exposed with a different dose of helium ions.
The effect of He-induced defects on the Raman spectrum of graphene is known from literature\cite{Buchheim2016,Shorubalko2016}, but for our analysis we explicitly do not rely on any a-priori knowledge of the system, i.e., we do not need to know beforehand which Raman bands will be altered by the irradiation and by what spatial pattern of the graphene has been irradiated. Instead, we use our clustering approach to identify the different types of Raman spectra present in the sample from which we infer the spatial distribution of He-irradiation doses and their effect on the graphene spectrum.
The sample under study consists of a free-standing graphene membrane (6 $\mu m$ diameter), suspended over a silicon nitride frame coated with Ti/Au (5~nm/40~nm). 
An illustration of the sample layout is presented in Fig.~\ref{fig:Raman}a. On this sample, a two-dimensional map containing 70$\times$70 spectra was acquired using a confocal Raman microscope (WITec alpha300 R) with a 532~nm excitation laser. A description of the sample preparation and Raman measurements is provided in Supplementary Note 8.\\

The Raman spectra were fed to the 28$\times$28 + UMAP (cos.) feature space construction method and split in 7 clusters using the GAL algorithm (see Supplementary Note 8 for more details). Figure~\ref{fig:Raman}b presents the partitioned feature space, containing several well-separated clusters. From this partitioning, we construct a two-dimensional map of the clusters to investigate their spatial distribution (see Fig.~\ref{fig:Raman}c). The plot shows that the extracted clusters match well the physical topology of the sample: Clusters 1-4 are located on the suspended graphene membrane, reproducing the four quadrants. Clusters 5-7 form concentric rings located at the edge of the boundary between the SiN/Ti/Au support and the hole and on the support itself.

Figure~\ref{fig:Raman}d shows the average spectrum obtained per cluster from the which the following characteristics can be evoked:
Cluster 1 shows a flat background, with pronounced peaks at 1585~cm$^{-1}$ and 2670~cm$^{-1}$. 
For Clusters 2 to 4 (corresponding to increasing He-dose), a peak at 1340~cm$^{-1}$ appears with steadily increasing intensity while the intensity of the peak at 2670~cm$^{-1}$, on the other hand, decreases. 
Cluster 5, located at the edge of the support possess all three above-mentioned peaks, while for Clusters 6 and 7, a broad fluorescence background originating from the gold is present and all graphene-related peaks drastically decrease in prominence. 
Interestingly, the four quadrants have only been identified as distinct clusters on the suspended part, but not on the substrate. This implies that the clustering algorithm identifies spectral changes upon irradiation as characteristic features for the freely suspended material, whereas the additional fluorescence background from the gold is a more characteristic attribute of the supported material than the variation between quadrants. Nevertheless, when inspecting Cluster 6 and 7, some sub structure is still visible, and performing a clustering on that subset may reveal additional structure.

The three observed peaks correspond to the well-known D-, G- and 2D-peak, and follow the behavior expected for progressive damage to graphene by He-irradiaton\cite{Buchheim2016,Shorubalko2016}. We would like to stress that our approach allowed to extract the increase of the D-peak and the decrease of the 2D-peak when introducing defects in graphene, without any before-hand knowledge of the system: neither the type of Raman spectra under consideration, nor where on the sample the He-irradiation occurred.

\section*{DISCUSSION}
In the synthetic data, the t-SNE and UMAP algorithms score equally well in reducing each measurement from a 784 dimensional space (28$\times$28) down to the 3 dimensional feature space. On the experimental datasets, however, UMAP tends to perform better. This difference emphasizes the need for labelled data which resembles as closely as possible the experimental data, as synthetic data may not capture all the experimental complexity. We note that UMAP has become the new state-of-the-art method for dimensionality reduction, surpassing t-SNE in several applications\cite{Becht2019,Diaz-Papkovich2019}. While t-SNE reproduces well the local structure of the data, UMAP reproduces both the local and large-scale structure\cite{McInnes2018}. Moreover, one could also investigate more advanced variants of UMAP\cite{McConville2019} that could lead to even higher FM indices. Along the same lines, the use of more sophisticated clustering algorithms involving convolutional neural networks that can directly be applied to the 28$\times$28 image merit additional research as some of them have proven to be highly accurate on the MNIST and other databases\cite{Ji2018}, despite their high computational cost. 

\section*{CONCLUSION}
In conclusion, we have introduced an optimized three-step workflow for the classification of univariate measurement data. The first two steps (feature space construction and partition algorithm) are based on an extensive benchmark of a wide range of novel and existing methods using 900 simulated datasets with known classes synthesized from experimental break junction traces. By doing so, we have identified specific combinations of feature space construction and partition algorithm yielding high accuracies, highlighting that a careful selection of the feature space construction and partition algorithm can significantly improve the classification results. We also provide guidelines for the estimation of the optimal number of clusters using a wide range of cluster validation indices. We show that our approach can readily be applied to various types of measurements such as MCBJ conductance-breaking traces, IV curves and Raman spectra, thereby splitting the dataset into statically relevant behaviors.

\section*{Data availability}
The experimental datasets used in this study are freely available online at https://doi.org/10.6084/m9.figshare.13258640. The generated datasets used for the benchmark shown in the main text are available at 10.6084/m9.figshare.13258595. The additional datasets generated for the benchmark are available from the corresponding author upon reasonable request.

\section*{Code availability}
The code used for this benchmark is freely available online at https://github.com/MickaelPerrin74/ClusteringBenchmark. In addition, we provide a graphical user interface for clustering data in a user-friendly fashion, containing all feature space construction methods and clustering algorithms used in this study. The code of this GUI is freely available online at 
https://github.com/MickaelPerrin74/DataClustering.

\section*{Acknowledgements}
 The authors would like to thank Dr. Ivan Shorubalko (Empa) for the help in developing the graphene membranes technology and for the He-FIB exposures (supported by Swiss National Science Foundation REquip 206021-133823). The authors would also like to thank Dr. Davide Stefani (Delft) for sharing with us the dataset recorded on OPE3\cite{Cabosart2019}.
M.P. acknowledges funding by the EMPAPOSTDOCS-II program which is financed by the European Unions Horizon 2020 research and innovation program under the Marie Sk\l{}odowska-Curie grant agreement number 754364. M.P. also acknowledges funding by the Swiss National Science Foundation (SNSF) under the Spark project no. 196795. This work was in part supported by the FET open project QuIET (no. 767187).

\section*{Author contributions}

M.P. performed the machine learning analysis, with input from all authors. J.O. and O.B. performed the Raman measurements. ME, JO, OB, MC, HvdZ, and MP discussed the data and wrote the manuscript. M.P. supervised the study.

\section*{Competing interests}
The authors declare no competing interests.

\section*{References}
\bibliography{references}

\end{document}